\documentclass{article}
\usepackage[english]{babel}
\usepackage[utf8]{inputenc}
\usepackage{spconf,ifpdf}
\usepackage{cite} 
\usepackage{url}
\usepackage[draft,bookmarks=false]{hyperref} 
\ifpdf
	\usepackage[pdftex]{graphicx}
	\graphicspath{{./figures/}}
\else
	\usepackage[dvips]{graphicx}
	\graphicspath{./figures/}
\fi
\usepackage{color}
\usepackage{pgf, tikz, pgfplots}

\usetikzlibrary{shapes, arrows, automata}
\usepackage{caption} 
\usepackage{subcaption} 
\usepackage{amsmath}
\usepackage{amsfonts, amssymb, amsthm}
\usepackage{mathrsfs}
\usepackage{upgreek}
\usepackage{algorithm,algpseudocode}
\usepackage{enumerate}
\usepackage{multirow}
\usepackage{rotating}
\usepackage{bm}
\usepackage{dsfont}
\usepackage{needspace}

\input{mySymbol.sty}

\newtheorem{assumption}{\hspace{0pt}\bf Assumption}
\newtheorem{lemma}{\hspace{0pt}\bf Lemma}
\newtheorem{proposition}{\hspace{0pt}\bf Proposition}

\newtheorem{theorem}{\hspace{0pt}\bf Theorem}

\newtheorem{definition}{\hspace{0pt}\bf Definition}

\newcommand{\QED}{\hfill\ensuremath{\blacksquare}}

\title{Stable and Transferable Wireless Resource Allocation Policies via Manifold Neural Networks}
{\name{Zhiyang Wang$^{\dagger}$ \qquad Luana Ruiz$^{\dagger}$ \qquad Mark Eisen$^{\star}$ \qquad Alejandro Ribeiro$^{\dagger}$}
\address{  ${^\dagger}$ Department of Electrical and Systems Engineering, University of Pennsylvania, PA \\
${^\star}$Intel Labs, Hillsboro, OR 
\thanks{Supported by NSF CCF 1717120, Theorinet Simons. } }
}

\begin{document}
%
\maketitle
%
\begin{abstract}
We consider the problem of resource allocation in large scale wireless networks. When contextualizing wireless network structures as graphs, we can model the limits of very large wireless systems as manifolds. To solve the problem in the machine learning framework, we propose the use of Manifold Neural Networks (MNNs) as a policy parametrization. In this work, we prove the stability of MNN resource allocation policies under the absolute perturbations to the Laplace-Beltrami operator of the manifold, representing system noise and dynamics present in wireless systems. These results establish the use of MNNs in achieving stable and transferable allocation policies for large scale wireless networks. We verify our results in numerical simulations that show superior performance relative to baseline methods.
\end{abstract}
%


\section{Introduction} \label{sec:intro}

With more devices and high-load applications deployed in wireless systems, allocating resources that mitigate interference and meet finite resource limitations across large networks has become an increasingly difficult challenge. While the resource allocation problem can be easily formulated as an optimization problem, due to non-convexity and large dimensionality, the exact solution is often non-tractable. Traditional heuristic methods have been used but require explicit model knowledge and large computation cost. 
In light of these difficulties, machine learning has become a valuable tool in tackling large-scale wireless resource allocation problems \cite{xu2019energy,sun2017learning,xu2017deep, eisen2020optimal,wang2021learning}. 

Modern machine learning techniques often involve training neural networks as a parametrization of the resource allocation policies. The specific parametric form imposed on the policy plays an important role in both the performance and the generalization. In particular, wireless systems require policies that are \emph{stable} to perturbations of the network states because (i) channel measurements are often noisy due to environmental factors and (ii) wireless networks change frequently in reality and it is impractical to retrain the neural network for each configuration. Graph neural networks (GNNs) \cite{gama2019convolutional} have been recently considered in many wireless resource allocation problems due to their low dimensionality and invariance to network topological structure \cite{eisen2020optimal, shen2020graph,chowdhury2021unfolding,wang2021learning}. While prior numerical results have been able to demonstrate transferability of GNN-based resource allocation policies \cite{eisen2020optimal, wang2021learning}, 
theoretical GNN stability results fail to capture this property when the size of graph is very large \cite{gama2020stability,zou2020graph}. 
Since our focus is on resource allocation problems over large-scale wireless networks, in this paper we take a different approach by approximating very large networks by continuous manifolds and modeling resource allocation policies in these networks via Manifold Neural Networks (MNNs) \cite{wang2021stability}.
Then, we prove the stability of such policies to formally establish their stability with respect to environmental/measurement noise, and demonstrate transferable performance across wireless networks with different scales. 


Specifically, wireless network interference patterns can be modeled as discrete graphs with the edges representing the dynamic interference channel states. 
As the number of network devices increases in the limit, the discrete graph can be represented as a continuous manifold structure $\ccalM$ \cite{belkin2008towards,calder2019improved}. We then consider the large scale resource allocation policy as processing inputs over a manifold and subsequently propose the Manifold Neural Network (MNN)---in effect representing GNNs on large graphs---as a suitable parametrization model. 
By modeling the system noise and dynamic changes as a perturbation to the Laplace-Beltrami operator of the underlying manifold $\ccalM$, we can analytically prove the stability of an MNN composed of frequency difference threshold (FDT) filters to such system perturbations. The learned MNN parametrization can moreover be transferred to both similar manifolds and to finitely sized graphs of increasing size \cite{levie2019transferability}. These results demonstrate the ability of MNNs to provide stable and transferable resource allocation policies for large scale wireless networks.

The rest of this paper is organized as follows. We formulate resource allocation problems over large scale graphs and manifolds (Section \ref{sec:resource}). We define MNNs as a parametrization of such resource allocation policies (Section \ref{sec:mnn}). In Section \ref{sec:properties}, we prove stability of MNN policies to absolute perturbations. In Section \ref{sec:sims}, we provide numerical analysis of GNNs in large scale wireless networks to validate the stability properties of MNNs and evaluate their transferability across networks.

\section{Optimal Resource Allocation} \label{sec:resource}


We consider a wireless network consisting of $m$ pairs of transmitters and receivers with each pair given a label $i \in \{1, 2, \hdots, m\}$. The channel link state can be characterized by a matrix $\bbS\in \reals^{m\times m}$ whose entries $s_{ij}:=[\bbS]_{ij}$ reflect the fading state between pairs $i$ and $j$. Note that the diagonal entries $s_{ii}$ reflect the direct channel between the transmitter and receiver in pair $i$ while off-diagonal entries $s_{ij}$ reflect an interference caused by transmitter $i$ at receiver $j$. We further consider the transmitter state $\bbx\in\reals^m$ with each entry $[\bbx]_i$ representing some state of transmitter $i$. 

The goal in resource allocation problems is to determine the optimal resource level $p_i$ for each transmitter $i$ under a given set of state observations $(\bbS, \bbx)$. In particular, we seek a mapping, or resource allocation policy, $\bbp(\bbS,\bbx):=[p_1;p_2;\hdots;p_m]$. For an instantaneous state and resource allocation, each pair $i$ experiences some level of performance $r_i(\bbp(\bbS,\bbx),\bbS,\bbx)$ (e.g. capacity, packet error rate). In fast-fading environments, the channel and transmitter states vary fast and randomly over time, and thus we optimize performance over the long-term distribution of performance measures obtained by the policy $\bbp(\bbS,\bbx)$. Due to the challenges inherent in optimizing an arbitrary function, the allocation policy $\bbp(\bbS,\bbx)$ is typically restricted to a vector-valued function family $\bm\phi(\bbH,\bbS,\bbx)$ parameterized by some parameter set $\bbH$. The optimal resource allocation policy is then formalized as determining the optimal policy parameter $\bbH^*$ as
\begin{align}
\label{eqn:opt}
    \bbH^* := &\argmax_{\bbH} \quad  \mathbb{E}_{\bbS, \bbx} \left[ \sum_{i=1}^m r_i\left( \bm{\phi}(\bbH,\bbS,\bbx), \bbS,\bbx\right)\right],\\
   \nonumber & \quad \text{s.t.}\qquad \quad \mathbb{E}_{\bbS, \bbx}\left [\sum_{i=1}^m \phi_i(\bbH,\bbS,\bbx)\right]\leq P_{\max}
  ,\\\nonumber& \quad \qquad \qquad {\phi}_i(\bbH,\bbS,\bbx)\in \{0,p_0\}, \quad i=1,..,m.
\end{align}
The optimal allocation policy $\bbphi(\bbH^*, \bbS,\bbx)$ is the one that maximizes the sum of performance measures under a constraint $P_{\max}$ on the total resource budget.


{Note that, in problem \eqref{eqn:opt}, $\bbS$ can be seen as the adjacency matrix of a very large graph $\bbG$ where each transmitter-receiver pair is a node and the connecting edges reflect the interference channels between pairs. Similarly, the transmitter states $\bbx$ can be seen as a graph signal on the nodes of $\bbG$.
In large scale wireless networks, where the number of transmitter/receiver pairs $m$ is very large, the full network channel state can instead be modeled as a continuum of interfering links between devices. In this paper, we model a large wireless system structure as a manifold $\ccalM$, i.e., the limit of graph $\bbG$ as $m \to \infty$.
}


Explicitly, we consider the channel state to be a smooth $d$-dimensional manifold $\ccalM$, which is a topological space that is locally Euclidean. The transmitter states are modeled as a manifold signal $f:\ccalM \rightarrow \reals$, which can be seen as the limit of the graph signal $\bbx$ when the network grows very large. Therefore, we can reformulate \eqref{eqn:opt} in the limit sense as
\begin{align}
\label{eqn:opt_manifold}
    \bbH^* := &\argmax_{\bbH}\; \mathbb{E}_{\ccalM, f}  \int\limits_{u\in\ccalM} r\left( \bm{\phi}(\bbH,\ccalM,f)(u), \ccalM,f(u)\right)\text{d}u,\\
   \nonumber & \quad \text{s.t.}\qquad  \mathbb{E}_{\ccalM, f}\int_{u\in\ccalM} \bm\phi(\bbH,\ccalM,f)(u)\text{d}u\leq P_{\max}
  ,\\& \nonumber\quad \qquad \qquad {\bm\phi}(\bbH,\ccalM, f)(u)\in [0,p_0], \quad u\in\ccalM.
\end{align}
{Observe in \eqref{eqn:opt_manifold} that the resource allocation is processed over a manifold $\ccalM$, while the utility and constraints are evaluated over the manifold rather than a discrete set of nodes.} {While continuous manifolds cannot be directly measured or observed in practice, the modeling of very large graphs as manifolds in \eqref{eqn:opt_manifold} is used in this paper as an analytical tool necessary for establishing the desired stability properties of policies $\bbphi(\bbH, \cdot, \cdot)$.} 

The challenge with problem \eqref{eqn:opt_manifold} is that the manifold is infinite-dimensional, so the policy parametrization has to be independent of the manifold dimension. This requirement is satisfied by a convolutional parametrization; thus, in the following we propose to parametrize $\bm{\phi}(\bbH,\ccalM,f)$ as a Manifold Neural Network (MNN). 

\subsection{Resource Allocation with Manifold Neural Networks}\label{sec:mnn}

In this paper we consider $d$-dimensional manifolds $\ccalM$ which are smooth and compact embedded submanifolds of Euclidean space. To each manifold $\ccalM$, a unique operator $\ccalL$ can be associated which characterizes how information propagates on $\ccalM$. This operator, called Laplace-Beltrami operator, is defined as $\ccalL f(u)=-\text{div}(\nabla f)(u)$, which is the divergence of the gradient of manifold signal $f$ in the local Euclidean space around the point $u \in \ccalM$. Akin to the Laplacian matrix in graphs, the Laplace-Beltrami operator measures the total variation of the manifold signal $f$.

The Laplace-Beltrami operator is a self-adjoint and positive-semidefinite operator. Therefore, it has a discrete, real and non-negative spectrum $\{\lambda_i,\bm\varphi_i\}_{i\in\naturals^+}$ which satisfies $\ccalL\bm\varphi_i = \lambda_i\bm\varphi_i$. Ordering the $\lambda_i$ in increasing order, i.e., $0 \leq \lambda_1 \leq \lambda_2 \leq \ldots$, it can be shown that $\lambda_i$ grows as $i^{2/d}$ where $d$ is the manifold dimension \cite{arendt2009weyl}. The eigenfunctions $\{\bm\varphi\}_{i\in\naturals^+}$ form an orthonormal basis of signals $f: \ccalM \to \reals$. Thus, a square-integrable signal $f$ can be represented as $f=\sum_{i=1}^\infty \langle f, \bm\varphi_i \rangle \bm\varphi_i$. 

Leveraging the eigendecomposition of $\ccalL$, the \textit{spectral convolution} of a manifold signal can be expressed as the filter 
\begin{equation}\label{eqn:operator}
\bbH(\ccalL) f:=\sum_{i=1}^\infty \sum_{k=0}^{K-1} h_k \lambda_i^k \langle f,\bm\varphi_i \rangle \bm\varphi_i,
\end{equation}
where $h_0, \ldots, h_{K-1}$ are coefficients that define a filter function $h(\lambda) = \sum_{k=0}^{K-1} h_k \lambda^k$ determining the amplification of the signal's spectral components based on their eigenvalues. 
From \eqref{eqn:operator}, we see that the manifold convolution only depends on the filter function and the Laplace-Beltrami operator eigenpairs. Hence, a manifold convolutional filter can be easily \textit{transferred} to other manifolds by replacing operator $\ccalL$. 

Given the convolutional filter in \eqref{eqn:operator}, Manifold Neural Networks (MNNs) are defined as a cascade of layers where each layer consists of a bank of manifold convolutional filters and a nonlinear activation function. Letting $\sigma_l$ denote the activation function at layer $l$, the $p$-th output feature of the $l$-th layer of a MNN can be written as
\begin{equation} \label{eqn:mnn}
    f_l^p = \sigma_l\left( \sum_{q=1}^{F_{l-1}} \bbH_l^{qp}(\ccalL)f_{l-1}^{q}\right)
\end{equation}
where, for $1\leq p\leq F_l$ and $1\leq q\leq F_{l-1}$, $\bbH_l^{qp}$ is the filter mapping the $q$-th feature from layer $l-1$ to the $p$-th feature of layer $l$. The output features of the last layer, given by $f_L^p$ for $1 \leq p \leq F_L$, are the MNN outputs $g^p$. The input features at the first layer, $f_0^q$, are the input data $f^q$ for $1\leq q\leq F_0$. 
Since in problem \eqref{eqn:opt_manifold} the transmitter states and the policy {are one-dimensional on the transmitter-receiver pairs}, when parametrizing $\bm{\phi}(\bbH,\ccalM,f)$ as a GNN we have $F_0=F_L=1$. Letting $g=g^1$ and $f=f^1$, we can thus represent the MNN \eqref{eqn:mnn} more succinctly as $g=\bm{\phi}(\bbH,\ccalM,f)$.

\section{Stability Analysis} \label{sec:properties}

In this paper, we analyze the stability of MNNs to absolute perturbations of the Laplace-Beltrami operator, which are presented in Definition \ref{defn:absolute_perturbations}.


\begin{definition}[Absolute perturbations] \label{defn:absolute_perturbations}
Let $\ccalL$ be the Laplace-Beltrami operator of a  manifold $\ccalM$. An absolute perturbation of $\ccalL$ is defined as 
\begin{equation}\label{eqn:perturb}
\ccalL'=\ccalL+\bbA,
\end{equation}
where the absolute perturbation operator $\bbA$ is symmetric.
\end{definition}
 
The absolute perturbation model introduced in Definition \ref{defn:absolute_perturbations} is a rather general perturbation model including many different types of perturbations. In particular, it can be used to approximate additive environmental noise in wireless fading channel states.

\subsection{Frequency difference threshold (FDT) filters}

The main challenge with absolute perturbations of $\ccalL$ is that they lead to perturbations of its spectrum. While these perturbations can be characterized individually, their cumulative effect is hard to measure because the spectrum is infinite-dimensional.
Nonetheless, this issue is alleviated by the fact that the eigenvalues of $\ccalL$ accumulate in a portion of the spectrum. This is formalized in Proposition \ref{prop:finite_num} which is a direct consequence of Weyl's law.



\begin{proposition}[\cite{arendt2009weyl}] \label{prop:finite_num}
Let $\ccalM$ be a $d$-dimensional embedded manifold with Laplace-Beltrami operator $\ccalL$, and let $\lambda_k$ denote the eigenvalues of $\ccalL$. Let $C_1$ denote an arbitrary constant, $C_d$ be the volume of the $d$-dimensional unit ball and $\text{Vol}(\ccalM)$ the volume of the manifold. For any $\alpha > 0$, there exists $N_1$ given by
\begin{equation}
    N_1=\lceil (\alpha d/C_1)^{d/(2-d)}(C_d \text{Vol}(\ccalM))^{2/(2-d)} \rceil
\end{equation}
such that, for all $k>N_1$, it holds that $$\lambda_{k+1}-\lambda_k\leq \alpha.$$
\end{proposition}

Proposition \ref{prop:finite_num} is important because it allows us to gather eigenvalues that are close enough, i.e., less than $\alpha$ apart, into a finite number of groups. This eigenvalue grouping, called the $\alpha$-separated spectrum, is introduced in Definition \ref{def:alpha-spectrum}. The manifold filters which achieve it, called Frequency Difference Threshold (FDT) filters, are presented in Definition \ref{def:alpha-filter}.


\begin{definition}[$\alpha$-separated spectrum]\label{def:alpha-spectrum}
The $\alpha$-separated spectrum of a Laplace-Beltrami operator $\ccalL$ is defined as the partition $\Lambda_1(\alpha) \cup \ldots\cup \Lambda_N(\alpha)$ such that, all $\lambda_i \in \Lambda_k(\alpha)$ and $\lambda_j \in \Lambda_l(\alpha)$, $k \neq l$, satisfy
\begin{align*}
|\lambda_i - \lambda_j| > \alpha \text{.}
\end{align*}
\end{definition}


\begin{definition}[$\alpha$-FDT filter]\label{def:alpha-filter}
The $\alpha$-frequency difference threshold ($\alpha$-FDT) filter is defined as a filter $\bbh(\ccalL)$ whose frequency response satisfies
\begin{equation} \label{eq:fdt-filter}
    |h(\lambda_i)-h(\lambda_j)|\leq \Delta_k \mbox{ for all } \lambda_i, \lambda_j \in \Lambda_k(\alpha) 
\end{equation}
with $\Delta_k\leq \Delta$ for $k=1, \ldots,N$.
\end{definition}

Eigenvalues belonging to different groups, i.e., $\lambda_i\in\Lambda_k(\alpha)$ and $\lambda_j\in\Lambda_l(\alpha)$ for $k\neq l$, are at least $\alpha$ apart from each other. Conversely, eigenvalues within the same group, i.e., $\lambda_i,\lambda_j\in\Lambda_k(\alpha)$, are always less than $\alpha$ apart. The $\alpha$-FDT filter achieves this spectrum separation by giving similar frequency responses---which differ by at most $\Delta_k$---for all $\lambda_i \in \Lambda_k(\alpha)$.  

\begin{figure}[t]
    \centering
 \includegraphics[width=0.4\textwidth]{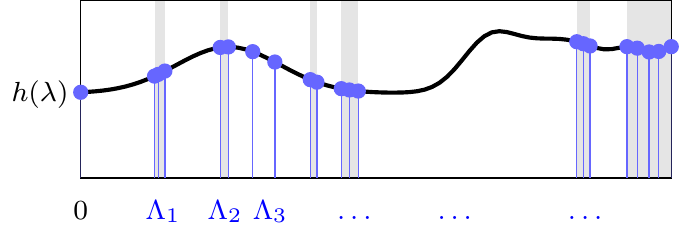}
    \caption{An $\alpha$-FDT filter that separates the spectrum of $\ccalL$ by grouping eigenvalues that are less than $\alpha$ apart.}
    \label{fig:filter-response}
\end{figure}

\subsection{Manifold Neural Network Stability}
As a consequence of spectrum separation, we can show that manifold neural networks composed of $\alpha$-FDT manifold filters are stable to the absolute perturbations to the Laplace-Beltrami operator as specified in Definition \ref{defn:absolute_perturbations}.
This is stated and proved in Theorem \ref{thm:stability_nn} under Assumptions \ref{ass:filter_function} and \ref{ass:activation}.

\begin{assumption}\label{ass:filter_function}
The filter function $h:\reals\rightarrow\reals$ is $B$- Lipschitz continuous and non-amplifying, i.e.,
\begin{equation}\label{eqn:filter_function}
    |h(a)-h(b)|\leq B|a-b|,\quad |h(a)|< 1.
\end{equation}
\end{assumption}

\begin{assumption}\label{ass:activation}
 The activation function $\sigma$ is normalized Lipschitz continous, i.e., $|\sigma(a)-\sigma(b)|\leq |a-b|$, with $\sigma(0)=0$.
\end{assumption}

\begin{theorem}[Manifold Neural network stability]\label{thm:stability_nn}
 Let $\ccalM$ be a manifold with Laplace-Beltrami operator $\ccalL$. Let $f$ be a manifold signal and $\bm\phi(\bbH,\ccalL,f)$ an $L$-layer manifold neural network on $\ccalM$ \eqref{eqn:mnn} with $F_0=F_L=1$ input and output features and $F_l=F,i=1,2,\hdots,L-1$ features per layer, and where the filters $\bbh(\ccalL)$ are $\alpha$-FDT filters with $\Delta=\pi\epsilon/(2\alpha-2\epsilon)$[cf. Definition \ref{def:alpha-filter}]. 
 Consider an absolute perturbation $\ccalL'=\ccalL + \bbA$ of the Laplace-Beltrami operator $\ccalL$ [cf. Definition \ref{defn:absolute_perturbations}] where $\|\bbA\| = \epsilon \leq \alpha$. 
 Then, under Assumptions \ref{ass:filter_function} and \ref{ass:activation} it holds:
 \begin{align}\label{eqn:stability_nn}
 \begin{split}
    \|\bm\phi(\bbH,\ccalL,f)-\bm\phi&(\bbH,\ccalL',f)\| \\
    &\leq LF^{L-1}\left(\frac{\pi N}{\alpha-\epsilon}+B\right)\epsilon \|f\|.
\end{split}
 \end{align}
where $N$ is the number of spectrum partitions. 
 \end{theorem}
 \begin{proof}
 See \cite{wang2021stable}.
 \end{proof}

Provided that Assumption \ref{ass:filter_function} and \ref{ass:activation} are satisfied, MNNs with $\alpha$-FDT filters are thus stable to absolute perturbations of the operator $\ccalL$. Note that these assumptions are reasonable because no constraint is put on the Lipschitz constant $B$, and most common activation functions, such as the ReLU, the modulus function and the sigmoid, satisfy the normalized Lipschitz continuity condition. From the bound in Theorem \ref{thm:stability_nn}, we see that MNN stability depends on the number of layers $L$ and the number of features per layer $F$, i.e., it is worse for deeper and wider MNNs. More importantly, we observe that MNNs have good stability if the Lipschitz constant $B$ is small and $\alpha$ is large. However, small $B$ and large $\alpha$ lead to less discriminative filters. While this reveals a stability-discriminability tradeoff, the presence of nonlinearities improves the discriminative power of MNNs because, akin to rectifiers, they spread some of the data's spectral content to parts of the spectrum where the next layer's FDT filters can discriminate it. {Therefore, convolutional neural network architectures provide a stable and transferable parametrization to resource allocation policies for large scale wireless networks.}



\section{Numerical experiments} \label{sec:sims}

In this section, we verify the transference and stability properties of the proposed {MNN resource allocation policies by numerically evaluating such properties with learned GNN-based policies on large graphs}.
While dropping $m$ transmitters randomly over a range of $\bba_i\in[-m,m]^2$, the paired receivers are dropped within  $\bbb_i\in[\bba_i+[-m/4,m/4]]^2$. When considering the large-scale pathloss gain and a random fast fading gain, the link state can be written as  $s_{ij}=\log( d_{ij}^{-2.2} h^f)$, where $d_{ij}$is the  distance between pair $i$ and $j$, while $h^f\sim  \text{Rayleigh}(2)$ is the random fading. The GNN is constructed with $L=10$ layers with a $K=5$ tap filter and a ReLu nonlinear activation function in each layer.

\begin{figure}[t]
    \centering
    \includegraphics[width=0.4\textwidth]{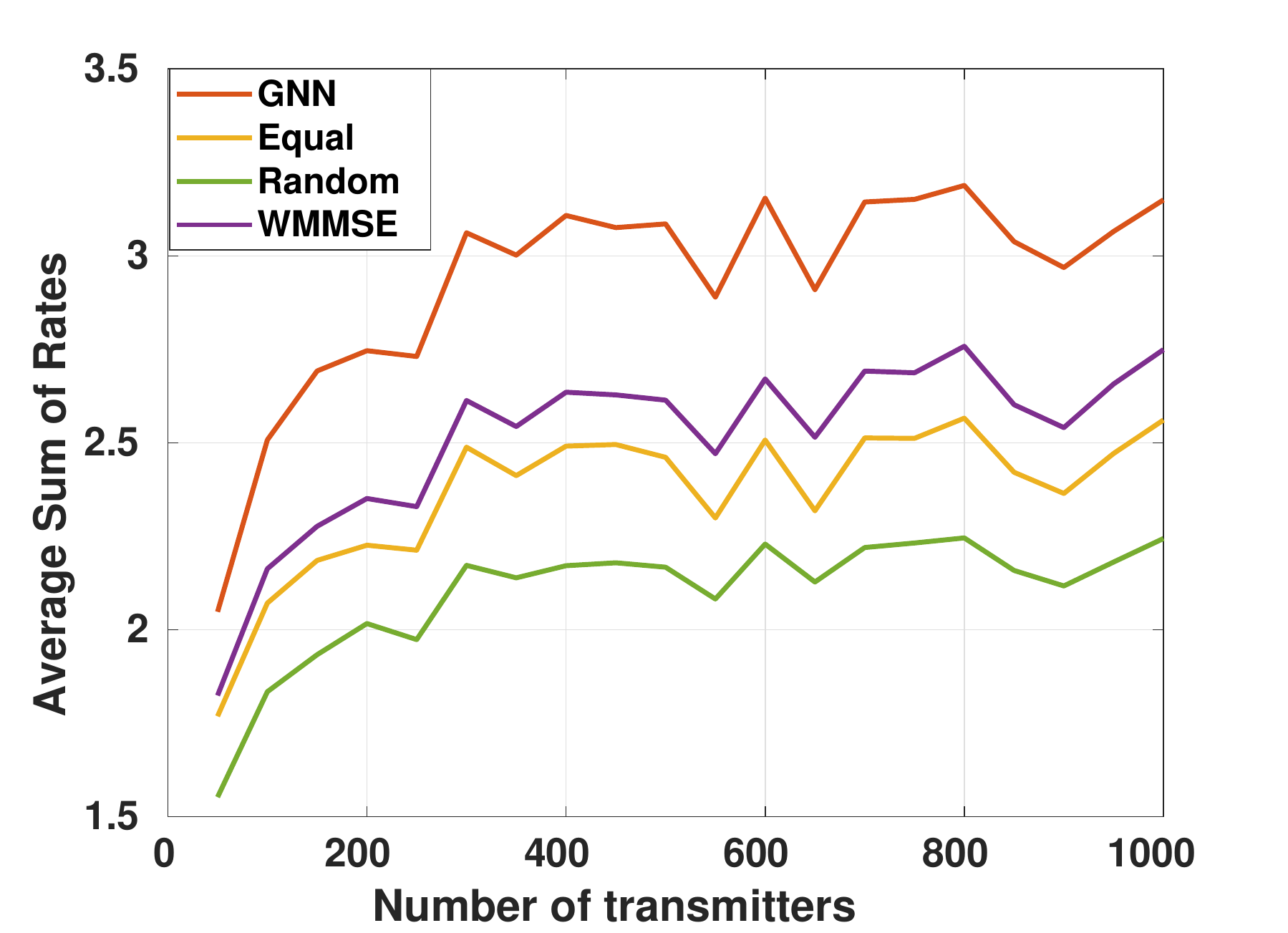}
    \caption{Sum-of-rate achieved by GNN trained on small network and execute on larger networks compared with other baseline methods.}
    \label{fig:sim_trans}
\end{figure}

When verifying the transferability of our proposed GNN methods, we compare with three existing baseline methods for solving this resource allocation problem. They are WMMSE \cite{sun2017learning}, equal resource allocation (i.e. assign $P_{max}/m$ to each transmitter) and random resource allocation (i.e. randomly select $P_{max}/p_0$ transmitters and assign $p_0$). We train a GNN policy on a network of size $m=50$ using unsupervised learning \cite{wang2021learning} and evaluate the trained policy on newly randomly generated wireless networks of larger size but the same overall network density. Observe the performance comparison shown in Figure \ref{fig:sim_trans} that the GNNs trained on smaller wireless networks can still outperform other methods on larger size networks, demonstrating transferable performance.

\begin{figure}[t]
    \centering
    \includegraphics[width=0.4\textwidth]{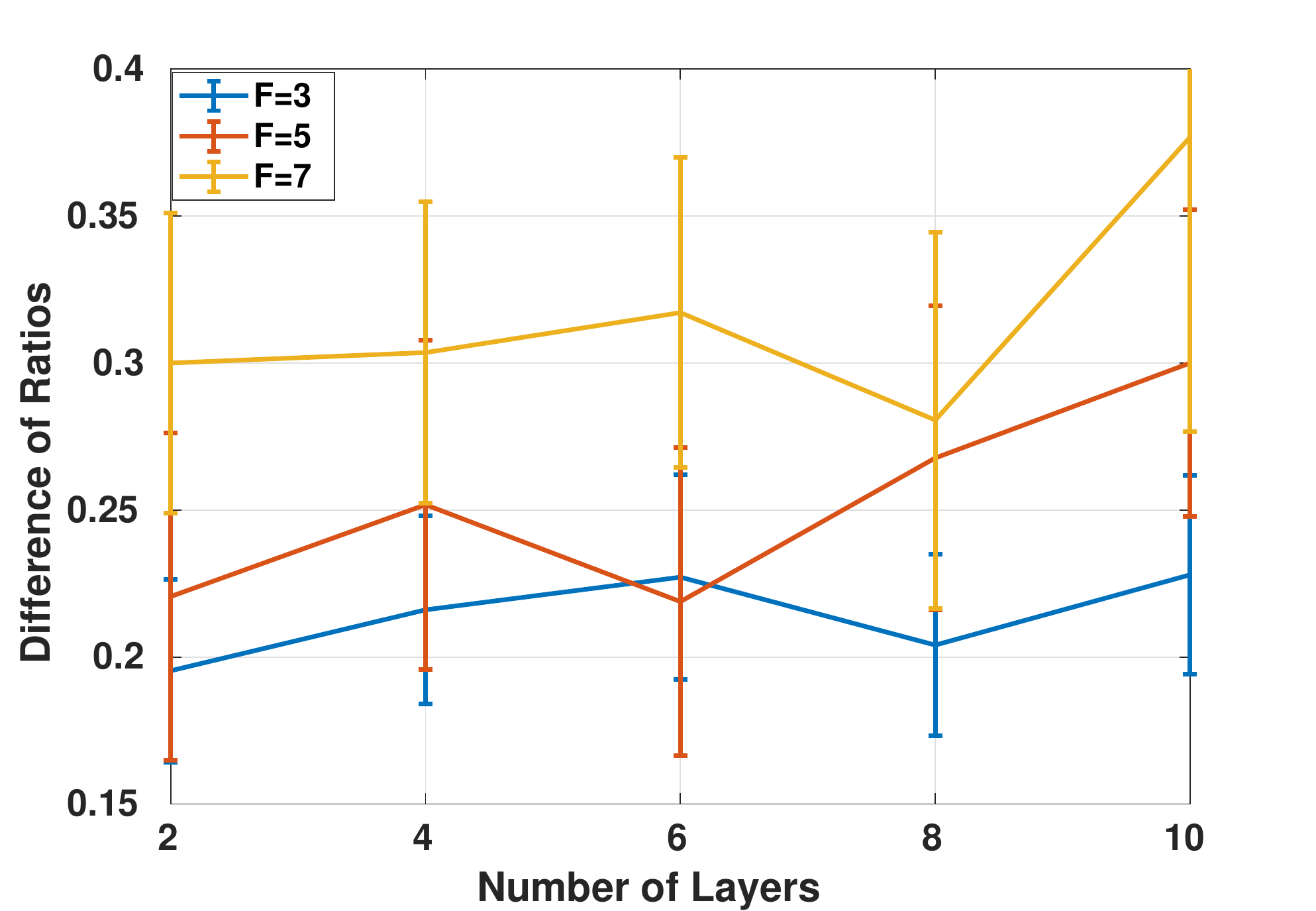}
    \caption{Difference between the sum-of-rate ratios on the original wireless network setting and the perturbed one.}
    \label{fig:sim_stability}
\end{figure}

To study the stability properties in large scale wireless system, we model environmental noise by adding a log-normal matrix to the original channel state $\bbS$ with $500$ pairs of transmitters and receivers. With the original trained GNN employed, we measure the stability by the difference of the ratios of the final sum-of-rate achieved by the GNN on the noisy channel states and that achieved by WMMSE. We can observe from Figure \ref{fig:sim_stability} that the difference increases with the number of layers and the number of filters per layer in the constructed GNN but is still generally small. This further validates the stability result in Theorem \ref{thm:stability_nn} of GNN-based resource allocation policies in large scale networks.


\section{Conclusions} \label{sec:conclusions}


In this paper, we have formulated the constrained resource allocation problem in large scale wireless systems. While the wireless network structure can be modeled as discrete graphs, large wireless networks tend to converge to a manifold structure as the number of network devices grow. We proposed a Manifold Neural Network (MNN) method for solving and analyzing resource allocation policies in large wireless networks. We proved that MNN is stable under the absolute perturbations to the Laplace-Beltrami operator of the manifold which in turn provides a stable and transferable allocation parametrization for large scale wireless networks. We verified results numerically with GNNs trained on large wireless network settings and compared performance with baseline methods.

 \appendix
 \section{Proof of Theorem 1 }
\begin{proof}
Begin with the definition of spectral convolution operators in \eqref{eqn:operator}, we can rewrite the norm difference between two outputs as:
\begin{align}
  \nonumber & \left\| \bbh(\ccalL) f - \bbh(\ccalL')f\right\| 
  \\
 & =\left\| \sum_{i=1}^\infty h(\lambda_{i}) \langle f, \bm\phi_i \rangle \bm\phi_i - \sum_{i=1}^\infty h(\lambda'_{i}) \langle f, \bm\phi'_i \rangle \bm\phi'_i \right\|. \label{eqn:diff}
\end{align}
We denote the index of partitions that contain a single eigenvalue as a set $\ccalK_s$ and the rest as a set $\ccalK_m$. We can decompose the filter function as $h(\lambda)=h^{(0)}(\lambda)+\sum_{l\in\ccalK_m}h^{(l)}(\lambda)$ with
\begin{align}
\label{eqn:h0}& h^{(0)}(\lambda) = \left\{ 
\begin{array}{cc} 
                h(\lambda)-\sum\limits_{l\in\ccalK_m}h(C_l)  &  \lambda\in[\Lambda_k(\alpha)]_{k\in\ccalK_s} \\
                0& \text{otherwise}  \\
                \end{array} \right. \text{ and}\\
\label{eqn:hl}& h^{(l)}(\lambda) = \left\{ 
\begin{array}{cc} 
                h(C_l) &  \lambda\in[\Lambda_k(\alpha)]_{k\in\ccalK_s} \\
                h(\lambda) & 
                \lambda\in\Lambda_l(\alpha)\\
                0 &
                \text{otherwise}  \\
                \end{array} \right.             
\end{align}
where $C_l$ is some constant in $\Lambda_l(\alpha)$. We can start by analyzing the output difference of $h^{(0)}(\lambda)$. With the triangle inequality, the norm difference can then be written as
\begin{align}
 & \nonumber \left\| \sum_{i=1}^\infty h^{(0)}(\lambda_{i}) \langle f, \bm\phi_i \rangle \bm\phi_i  -  h^{(0)}(\lambda'_i )  \langle f, \bm\phi'_i \rangle \bm\phi'_i \right\|  \\
 &\nonumber  =\Bigg\|\sum_{i=1}^\infty  h^{(0)}(\lambda_{i}) \langle f, \bm\phi_i \rangle \bm\phi_i -  h^{(0)}(\lambda_{i}) \langle f, \bm\phi'_i \rangle \bm\phi'_i
 \\& \qquad\qquad  + h^{(0)}(\lambda_{i}) \langle f, \bm\phi'_i \rangle \bm\phi'_i- h^{(0)}(\lambda'_{i}) \langle f, \bm\phi'_i \rangle \bm\phi'_i \Bigg\| \\
  & \nonumber \leq \left\|\sum_{i=1}^\infty  h^{(0)}(\lambda_i)  \langle f, \bm\phi_i \rangle \bm\phi_i -  h^{(0)}(\lambda_i ) \langle f, \bm\phi'_i \rangle \bm\phi'_i\right\| \\&\qquad  +\left\|\sum_{i =1}^\infty h^{(0)}(\lambda_{i}) \langle f, \bm\phi'_i \rangle \bm\phi'_i -  h^{(0)}(\lambda'_{i}) \langle f, \bm\phi'_i \rangle \bm\phi'_i \right\| \\
 &\nonumber \leq \Bigg\| \sum_{i=1}^\infty  h^{(0)}(\lambda_i)( \langle f, \bm\phi_i \rangle\bm\phi_i-\langle f, \bm\phi_i \rangle\bm\phi'_i +\langle f, \bm\phi_i \rangle\bm\phi'_i \\
 & \quad - \langle f, \bm\phi'_i \rangle \bm\phi'_i ) \Bigg\| +\Bigg\|\sum_{i =1}^\infty  (h^{(0)}(\lambda_i ) -h^{(0)}(\lambda'_i) ) \langle f, \bm\phi'_i \rangle \bm\phi'_i  \Bigg\| \\
& \nonumber \leq \left\| \sum_{i=1}^\infty h^{(0)}(\lambda_i )\langle f, \bm\phi_i \rangle (\bm\phi_i - \bm\phi'_i ) \right\| \\ \nonumber&\qquad\qquad  + \Bigg\|  \sum_{i =1}^\infty  h^{(0)}(\lambda_i )\langle f, \bm\phi_i - \bm\phi'_i  \rangle \bm\phi'_i \Bigg\|\\& \qquad\qquad \quad+ \left\|\sum_{i=1}^\infty  (h^{(0)}(\lambda_i ) -h^{(0)}(\lambda'_i) ) \langle f, \bm\phi'_i \rangle \bm\phi'_i  \right\| \label{eqn:3}
\end{align}
Now we need to include two important lemmas to analyze the influence on eigenvalues and eigenfunctions caused by the perturbation.
\begin{lemma}\label{lem:eigenvalue_absolute}[Weyl's Theorem]
The eigenvalues of LB operators $\ccalL$ and perturbed $\ccalL'=\ccalL+\bbA$ satisfy
\begin{equation}
|\lambda_i-\lambda'_i|\leq \|\bbA\|, \text{ for all }i=1,2\hdots
\end{equation}
\end{lemma}
\begin{proof}[Proof of Lemma \ref{lem:eigenvalue_absolute}]
The minimax principle asserts that
\begin{align}
   \nonumber \lambda_i(\ccalL)&=\max_{codim T = i-1}\lambda[\ccalL, T]\\ &=\max_{codim T \leq i-1} \min_{u\in T, \|u\|=1} \langle \ccalL u, u \rangle.
\end{align}
Then for any $1\leq k $, we have
\begin{align}
    \lambda_i(\ccalL') &=\max_{codim T\leq i-1} \min_{ u\in T, \|u\|=1}  \langle (\ccalL+\bbA) u, u\rangle \\
    & = \max_{codim T\leq i-1} \min_{ u\in T, \|u\|=1} \left( \langle  (\ccalL  u, u\rangle + \langle \bbA u, u\rangle   \right)\\
    & \geq \max_{codim T\leq i-1} \min_{  u\in T, \|u\|=1}  \left\langle  \ccalL  u, u\rangle   + \lambda_1(\bbA) \right)\\
    & = \lambda_1(\bbA)+ \max_{codim T\leq i-1} \min_{  u\in T, \|u\|=1 } \langle  \ccalS  u, u\rangle  \\
    & = \lambda_k(\ccalS)+\lambda_1(\bbA).
\end{align}
Similarly, we can have $\lambda_i(\ccalL') \leq \lambda_i(\ccalL)+ \max_k\lambda_k(\bbA)$. This leads to $\lambda_1(\bbA)\leq \lambda_i(\ccalS + \bbA )-\lambda_k(\ccalS) \leq \max_k\lambda_k(\bbA)$. This leads to the conclusion that:
\begin{equation}
    |\lambda'_i-\lambda_i|\leq \|\bbA\|.
\end{equation}
\end{proof}

To measure the difference of eigenfunctions, we introduce the Davis-Kahan $\sin\theta$ theorem as follows.
\begin{lemma}[Davis-Kahan $\sin\theta$ Theorem]\label{lem:davis-kahan}
Suppose the spectra of operators $\ccalL$ and $\ccalL'$ are partitioned as $\sigma\bigcup\Sigma$ and $\omega\bigcup \Omega$ respectively, with $\sigma\bigcap \Sigma=\emptyset$ and $\omega\bigcap\Omega=\emptyset$. Then we have
\begin{equation}
\|E_\ccalL(\sigma)-E_{\ccalL'}(\omega)\|\leq \frac{\pi}{2}\frac{\|(\ccalL'-\ccalL)E_\ccalL(\sigma)\|}{d}\leq \frac{\pi}{2}\frac{\|\ccalL'-\ccalL\|}{d},
\end{equation}
where $d$ satisfies $\min_{x\in\sigma,y\in\Omega}|x-y|\geq d$ and $\min_{x\in\Sigma,y\in\omega}|x-y|\geq d$.
\end{lemma}
\begin{proof}[Proof of Lemma \ref{lem:davis-kahan}] 
See \cite{seelmann2014notes}.
\end{proof}
For the first term in \eqref{eqn:3}, we employ Lemma \ref{lem:davis-kahan} and therefore we have $\sigma=\lambda_i$ and $\omega=\lambda'_i$, for $\lambda_i\in [\Lambda_k(\alpha)]_{k\in\ccalK_s}$ we can have
\begin{align}
\left\| \bm\phi_i -\bm\phi'_i \right\| \leq \frac{\pi}{2} \frac{\|\bbA\|}{\alpha-\epsilon}= \frac{\pi}{2} \frac{\epsilon}{\alpha-\epsilon}.
\end{align}
Here $d$ can be seen as $d=\min_{\lambda_i\in\Lambda_k(\alpha),\lambda_j\in\Lambda_l(\alpha),k\neq l}|\lambda_i-\lambda_j'|$. Combined with the fact that $|\lambda_i-\lambda_j|>\alpha$ and $|\lambda_i-\lambda_i'|\leq \epsilon$ for all $\lambda_i\in\Lambda_k(\alpha),\lambda_j\in\Lambda_l(\alpha),k\neq l$, we have $d\geq \alpha-\epsilon$. With Cauchy-Schwartz inequality, we have the first term in \eqref{eqn:3} bounded as
\begin{align}
&\nonumber\left\| \sum_{i=1}^\infty h^{(0)}(\lambda_i )\langle f, \bm\phi_i \rangle (\bm\phi_i - \bm\phi'_i ) \right\|\\
& \leq \sum_{i=1}^\infty |h^{(0)}(\lambda_i)| | \langle f, \bm\phi_i \rangle | \left\|\bm\phi_i-\bm\phi'_i \right\| \leq  \frac{N_s\pi\epsilon}{2(\alpha-\epsilon)}  \|f\|.
\end{align}
The second term in \eqref{eqn:3} is bounded as
\begin{align}
 &\nonumber \left\|  \sum_{i =1}^\infty  h^{(0)}(\lambda_i )\langle f, \bm\phi_i - \bm\phi'_i  \rangle \bm\phi'_i \right\| \\
 &\leq   \sum_{i =1}^\infty |h^{(0)}(\lambda_i)| \|\bm\phi_i - \bm\phi'_i \| \|f\| \leq   \frac{N_s \pi\epsilon}{2(\alpha-\epsilon)}  \|f\|.
\end{align}
These two bounds are obtained by noting that $|h^{(0)}(\lambda)|<1$ and $h^{(0)}(\lambda)=0$ for $\lambda\in[\Lambda_k(\alpha)]_{k\in\ccalK_m}$. The number of eigenvalues within $[\Lambda_k(\alpha)]_{k\in\ccalK_s}$ is denoted as $N_s$. The third term in \eqref{eqn:3} can be bounded by the Lipschitz continuity of $h$ combined with Lemma \ref{lem:eigenvalue_absolute}.
\begin{align}
\nonumber  \Bigg\|\sum_{i=1}^\infty  &(h^{(0)}(\lambda_i ) -h^{(0)}(\lambda'_i) ) \langle f, \bm\phi'_i \rangle \bm\phi'_i  \Bigg\|^2 
  \\ \nonumber & \leq \sum_{i=1}^\infty | h^{(0)}(\lambda_{i}) -h^{(0)}(\lambda'_i) |^2 |\langle f, \bm\phi'_i \rangle|^2 \\
  &\leq \sum_{i =1}^\infty B^2 |\lambda_i - \lambda'_i |^2 |\langle f, \bm\phi'_i \rangle|^2 \leq  B^2 \epsilon^2 \|f\|^2.
\end{align}

Then we need to analyze the output difference of $h^{(l)}(\lambda)$, we can bound this as
\begin{align}
    \nonumber &\left\| \bbh^{(l)}(\ccalL)f -\bbh^{(l)}(\ccalL')f \right\| 
    \\& \leq \left\| (h(C_l)+\Delta)f -(h(C_l)-\Delta)f\right\| \leq 2\Delta\|f\|,
\end{align}
where $\bbh^{(l)}(\ccalL)$ and $\bbh^{(l)}(\ccalL')$ are manifold filters with filter function $h^{(l)}(\lambda)$ on the LB operators $\ccalL$ and $\ccalL'$ respectively.
Combining the filter functions, we can write
\begin{align}
\label{eqn:sta-filter-alpha}
   \nonumber &\|\bbh(\ccalL)f-\bbh(\ccalL')f\|=\\&
    \left\|\bbh^{(0)}(\ccalL)f +\sum_{l\in\ccalK_m}\bbh^{(l)}(\ccalL)f - \bbh^{(0)}(\ccalL')f - \sum_{l\in\ccalK_m} \bbh^{(l)}(\ccalL')f \right\|\\
    &\leq \|\bbh^{(0)}(\ccalL)f-\bbh^{(0)}(\ccalL')f\|+\sum_{l\in\ccalK_m}\|\bbh^{(l)}(\ccalL)f-\bbh^{(l)}(\ccalL')f\|\\
    &\leq \frac{N_s\pi\epsilon}{\alpha-\epsilon}\|f\| + B\epsilon\|f\| +2(N-N_s)\Delta\|f\|.
\end{align}
With $\Delta$ set as $\frac{\pi\epsilon}{2(\alpha-\epsilon)}$, the bound became
\begin{align}
\label{eqn:sta-filter-alpha}
   \|\bbh(\ccalL)f-\bbh(\ccalL')f\|\leq \frac{N\pi\epsilon}{\alpha-\epsilon}\|f\| + B\epsilon\|f\|.
\end{align}
We can extend the stability result to the MNN. To bound the output difference $\|\bby-\bby'\|$, we need to write in the form of features of the final layer
 \begin{equation}
 \|\bm\phi(\bbH,\ccalL,f)-\bm\phi(\bbH,\ccalL',f)\| =\sum_{q=1}^{F_L} \| f_L^q - f_L^{'q}\|.
 \end{equation}
The output signal of layer $l$ of MNN $\bbPhi(\bbH,\ccalL, f)$ can be written as
\begin{equation}
 f_l^p = \sigma\left( \sum_{q=1}^{F_{l-1}} \bbh_l^{pq}(\ccalL) f_{l-1}^q\right).
\end{equation}
Similarly, for the perturbed $\ccalL'$ the corresponding MNN is $\bbPhi(\bbH,\ccalL',f)$ the output signal can be written as
 \begin{equation}
 f_l^{'p} = \sigma\left( \sum_{q=1}^{F_{l-1}} \bbH_l^{pq}(\ccalL') f_{l-1}^{'q}\right).
 \end{equation}
The difference therefore becomes
 \begin{align}
 &\nonumber\| f_l^p - f_l^{'p} \| \\& =\left\|  \sigma\left( \sum_{q=1}^{F_{l-1}} \bbH_l^{pq}(\ccalL) f_{l-1}^q\right) -  \sigma\left( \sum_{q=1}^{F_{l-1}} \bbH_l^{pq}(\ccalL') f_{l-1}^{'q}\right) \right\|.   
 \end{align}
With the assumption that $\sigma$ is normalized Lipschitz, we have
 \begin{align}
  &\nonumber\| f_l^p - f_l^{'p} \| \\&\qquad\qquad \leq \left\| \sum_{q=1}^{F_{l-1}}  \bbH_l^{pq}(\ccalL) f_{l-1}^q - \bbH_l^{pq}(\ccalL') f_{l-1}^{'q}  \right\| \\& \qquad\qquad \leq \sum_{q=1}^{F_{l-1}} \left\|  \bbH_l^{pq}(\ccalL) f_{l-1}^q - \bbH_l^{pq}(\ccalL') f_{l-1}^{'q} \right\|.
 \end{align}
By adding and subtracting $\bbH_l^{pq}(\ccalL') f_{l-1}^{q}$ from each term, combined with the triangle inequality we can get
 \begin{align}
 & \nonumber \left\|  \bbH_l^{pq}(\ccalL) f_{l-1}^q - \bbH_l^{pq}(\ccalL') f_{l-1}^{'q} \right\| \\\nonumber &\quad \leq \left\|  \bbH_l^{pq}(\ccalL) f_{l-1}^q - \bbH_l^{pq}(\ccalL') f_{l-1}^{q} \right\| \\&\qquad \qquad \qquad + \left\| \bbH_l^{pq}(\ccalL') f_{l-1}^q - \bbH_l^{pq}(\ccalL') f_{l-1}^{'q} \right\|
 \end{align}
The first term can be bounded with \eqref{eqn:sta-filter-alpha} for absolute perturbations. The second term can be decomposed by Cauchy-Schwartz inequality and non-amplifying of the filter functions as
 \begin{align}
 \left\| f_{l}^p - f_l^{'p} \right\| \leq \sum_{q=1}^{F_{l-1}} C_{per} \epsilon \| f_{l-1}^q\| + \sum_{q=1}^{F_{l-1}} \| f_{l-1}^q - f_{l-1}^{'q} \|,
 \end{align}
where we use $C_{per}$ to represent the terms in \eqref{eqn:sta-filter-alpha} for simplicity. To solve this recursion, we need to compute the bound for $\|f_l^p\|$. By normalized Lipschitz continuity of $\sigma$ and the fact that $\sigma(0)=0$, we can get
 \begin{align}
 \nonumber &\| f_l^p \|\leq \left\| \sum_{q=1}^{F_{l-1}} \bbh_l^{pq}(\ccalL) f_{l-1}^{q}  \right\| \leq  \sum_{q=1}^{F_{l-1}}  \left\| \bbh_l^{pq}(\ccalL)\right\|  \|f_{l-1}^{q}  \| \\
 &\qquad \leq   \sum_{q=1}^{F_{l-1}}   \|f_{l-1}^{q}  \| \leq \prod\limits_{l'=1}^{l-1} F_{l'} \sum_{q=1}^{F_0}\| f^q \|.
 \end{align}
 Insert this conclusion back to solve the recursion, we can get
 \begin{align}
 \left\| f_{l}^p - f_l^{'p} \right\| \leq l C_{per}\epsilon \left( \prod\limits_{l'=1}^{l-1} F_{l'} \right) \sum_{q=1}^{F_0} \|f^q\|.
 \end{align}
 Replace $l$ with $L$ we can obtain
 \begin{align}
 &\nonumber \|\bm\phi(\bbH,\ccalL,f) - \bm\phi(\bbH,\ccalL',f)\| \\
 &\qquad \qquad \leq \sum_{q=1}^{F_L} \left( L C_{per}\epsilon \left( \prod\limits_{l'=1}^{L-1} F_{l'} \right) \sum_{q=1}^{F_0} \|f^q\| \right).
 \end{align}
 With $F_0=F_L=1$ and $F_l=F$ for $1\leq l\leq L-1$, then we have
  \begin{align}
 \|\bm\phi(\bbH,\ccalL,f) - \bm\phi(\bbH,\ccalL',f)\| \leq LF^{L-1} C_{per}\epsilon \|f\|,
 \end{align}
 with $C_{per}=\frac{N\pi}{\alpha-\epsilon} +B$ as the terms in \eqref{eqn:sta-filter-alpha}.
\end{proof}

\bibliographystyle{IEEEtran}
\bibliography{references}

\begin{thebibliography}{10}
\providecommand{\url}[1]{#1}
\csname url@samestyle\endcsname
\providecommand{\newblock}{\relax}
\providecommand{\bibinfo}[2]{#2}
\providecommand{\BIBentrySTDinterwordspacing}{\spaceskip=0pt\relax}
\providecommand{\BIBentryALTinterwordstretchfactor}{4}
\providecommand{\BIBentryALTinterwordspacing}{\spaceskip=\fontdimen2\font plus
\BIBentryALTinterwordstretchfactor\fontdimen3\font minus
  \fontdimen4\font\relax}
\providecommand{\BIBforeignlanguage}[2]{{%
\expandafter\ifx\csname l@#1\endcsname\relax
\typeout{** WARNING: IEEEtran.bst: No hyphenation pattern has been}%
\typeout{** loaded for the language `#1'. Using the pattern for}%
\typeout{** the default language instead.}%
\else
\language=\csname l@#1\endcsname
\fi
#2}}
\providecommand{\BIBdecl}{\relax}
\BIBdecl

\bibitem{xu2019energy}
D.~Xu, X.~Che, C.~Wu, S.~Zhang, S.~Xu, and S.~Cao, ``Energy-efficient
  subchannel and power allocation for hetnets based on convolutional neural
  network,'' \emph{arXiv preprint arXiv:1903.00165}, 2019.

\bibitem{sun2017learning}
H.~Sun, X.~Chen, Q.~Shi, M.~Hong, X.~Fu, and N.~D. Sidiropoulos, ``Learning to
  {optimize}: Training deep neural networks for wireless resource management,''
  \emph{IEEE Transactions on Signal Processing}, vol.~66, no.~20, pp.
  5438--5453, 2018.

\bibitem{xu2017deep}
Z.~Xu, Y.~Wang, J.~Tang, J.~Wang, and M.~C. Gursoy, ``A deep reinforcement
  learning based framework for power-efficient resource allocation in cloud
  rans,'' in \emph{2017 IEEE International Conference on Communications
  (ICC)}.\hskip 1em plus 0.5em minus 0.4em\relax IEEE, 2017, pp. 1--6.

\bibitem{eisen2020optimal}
M.~Eisen and A.~Ribeiro, ``Optimal wireless resource allocation with random
  edge graph neural networks,'' \emph{IEEE Transactions on Signal Processing},
  vol.~68, pp. 2977--2991, 2020.

\bibitem{wang2021learning}
Z.~Wang, M.~Eisen, and A.~Ribeiro, ``Learning decentralized wireless resource
  allocations with graph neural networks,'' \emph{arXiv preprint
  arXiv:2107.01489}, 2021.

\bibitem{gama2019convolutional}
F.~Gama, A.~G. Marques, G.~Leus, and A.~Ribeiro, ``Convolutional neural network
  architectures for signals supported on graphs,'' \emph{IEEE Transactions on
  Signal Processing}, vol.~67, no.~4, pp. 1034--1049, 2019.

\bibitem{shen2020graph}
Y.~Shen, Y.~Shi, J.~Zhang, and K.~B. Letaief, ``Graph neural networks for
  scalable radio resource management: Architecture design and theoretical
  analysis,'' \emph{arXiv preprint arXiv:2007.07632}, 2020.

\bibitem{chowdhury2021unfolding}
A.~Chowdhury, G.~Verma, C.~Rao, A.~Swami, and S.~Segarra, ``Unfolding wmmse
  using graph neural networks for efficient power allocation,'' \emph{IEEE
  Transactions on Wireless Communications}, 2021.

\bibitem{gama2020stability}
F.~Gama, J.~Bruna, and A.~Ribeiro, ``Stability properties of graph neural
  networks,'' \emph{IEEE Transactions on Signal Processing}, vol.~68, pp.
  5680--5695, 2020.

\bibitem{zou2020graph}
D.~Zou and G.~Lerman, ``Graph convolutional neural networks via scattering,''
  \emph{Applied and Computational Harmonic Analysis}, vol.~49, no.~3, pp.
  1046--1074, 2020.

\bibitem{wang2021stability}
Z.~Wang, L.~Ruiz, and A.~Ribeiro, ``Stability of neural networks on riemannian
  manifolds,'' \emph{arXiv preprint arXiv:2103.02663}, 2021.

\bibitem{belkin2008towards}
M.~Belkin and P.~Niyogi, ``Towards a theoretical foundation for laplacian-based
  manifold methods,'' \emph{Journal of Computer and System Sciences}, vol.~74,
  no.~8, pp. 1289--1308, 2008.

\bibitem{calder2019improved}
J.~Calder and N.~G. Trillos, ``Improved spectral convergence rates for graph
  laplacians on epsilon-graphs and k-nn graphs,'' \emph{arXiv preprint
  arXiv:1910.13476}, 2019.

\bibitem{levie2019transferability}
R.~Levie, M.~M. Bronstein, and G.~Kutyniok, ``Transferability of spectral graph
  convolutional neural networks,'' \emph{arXiv preprint arXiv:1907.12972},
  2019.

\bibitem{arendt2009weyl}
W.~Arendt, R.~Nittka, W.~Peter, and F.~Steiner, ``Weyl’s law: Spectral
  properties of the laplacian in mathematics and physics,'' \emph{Mathematical
  analysis of evolution, information, and complexity}, pp. 1--71, 2009.

\bibitem{wang2021stable}
\BIBentryALTinterwordspacing
``Stable and transferable wireless resource allocation policies via manifold
  neural networks.'' [Online]. Available:
  \url{https://zhiyangw.com/Papers/Stable-icassp2022.pdf}
\BIBentrySTDinterwordspacing

\bibitem{seelmann2014notes}
A.~Seelmann, ``Notes on the $sin 2\theta$ theorem,'' \emph{Integral Equations
  and Operator Theory}, vol.~79, no.~4, pp. 579--597, 2014.

\end{thebibliography}

\end{document}